\documentclass[twocolumn,prd,showpacs,preprintnumbers,amsmath,amssymb]{revtex4}

\usepackage{amsmath,amssymb,amsthm}

\usepackage{epsfig}

\usepackage{graphicx}

\begin{document}

\title{Protecting quantum coherence of two-level atoms from vacuum fluctuations of electromagnetic field }
\author{  Xiaobao Liu, Zehua Tian, Jieci Wang and Jiliang Jing\footnote{Corresponding author, Email: jljing@hunn.edu.cn}  }
\affiliation{Department of Physics, Key Laboratory of Low Dimensional Quantum Structures and Quantum Control of Ministry of Education, and Synergetic Innovation Center for Quantum Effects and Applications, Hunan Normal University, Changsha, Hunan 410081, P. R. China
}

\begin{abstract}
In the framework of open quantum systems, we study the dynamics of a static polarizable two-level atom interacting with a bath of fluctuating vacuum electromagnetic
field and explore under which conditions the coherence of the open quantum system is unaffected by the environment totally.
For both a single-qubit and two-qubit systems,  we find that the quantum coherence can not be protected from noise when the atom interacts with a non-boundary electromagnetic field. However, with the presence of a boundary, the dynamical conditions for the insusceptible of quantum coherence are fulfilled only when the atom is close to the boundary and is transversely polarizable. Otherwise, the quantum coherence can only be protected in some degree in other polarizable direction.

\end{abstract}


\maketitle

\section{Introduction}
Quantum coherence is an important concept in quantum theory. The superposition principle of quantum states summarize the characteristic features of quantum systems from their classical counterparts, which marks the coherence is a fundamental aspect of quantum physics~\cite{Leggett}. As the technological progress in the last few decades,  quantum coherence becomes a powerful resource in various aspects, such as low-temperature thermodynamics~\cite{Narasimhachar,Lostaglio1,berg,Horodecki,Lostagio}, quantum metrology~\cite{Giovannetti,Demkowicz} and solid-state physics~\cite{Deveaud,-M}. Besides, the interference phenomena plays a key role in  quantum optics experiments~\cite{Glauber,Scully,Albrecht,Walls}, as well as a series of quantum information processing tasks~\cite{Nielsen}. Conventionally, quantum coherence is regarded as a physical resource which is associated with the capability of a quantum state to exhibit quantum interference phenomena. Recently, Baumgratz \emph{et al.} proposed some rigorous measurements to quantify coherence, such as \textit{$l_1$} norm and relative entropy of coherence~\cite{Baumgratz}, which are distance-based measurements for such resources.

On the other hand, a realistic quantum system should be regarded as an open system due to the interaction between the system and its surrounding environment. For example, a two-level atom bathing in a vacuum fluctuations of electromagnetic field would suffers from decoherence, in which quantum coherence would reduced due to the interaction between the atom and the environment. Considering that the coherence is not only a fundamental aspect of quantum physics but also a crucial resource in quantum information technology, it is essential to find strategies to protect it. Most recently, Bromley \emph{et al.} found that under some suitable dynamic conditions, the quantum coherence is efficiently preserved from noise during the entire evolution~\cite{thomas}.

In this paper we consider a static two-level atom which couples with the bath of fluctuating vacuum electromagnetic field. Our goal is to protect the quantum coherence by setting a reflecting boundary in the electromagnetic field, which would changes the vacuum fluctuations. We first calculate the evolution of the \textit{$l_1$} norm and relative entropy of coherence for a single-qubit system. Two different cases, the electromagnetic field without or with the presence of a reflecting boundary, are considered. We are going to analyze under which conditions the coherence is totally unaffected by the vacuum fluctuations.
We find that in an unbounded space, the quantum coherence can not be frozen during the whole evolution, which is due to the interaction between the two-level atom and the electromagnetic field. However, with the presence of a boundary, the quantum coherence can be frozen when the atom is close to the boundary and is transversely polarizable, which means that the quantum coherence can be protected efficiently. We then study the quantum coherence of a two-qubit system with maximally mixed marginals ~\cite{Horodecki1,Horodecki2} and find that the result is similar to that of the single-qubit case.

The outline of the paper is as follows. In section II, we briefly introduce the methods to quantify coherence and discuss the dynamics of open quantum systems for a static polarizable two-level atom coupling a bath of fluctuating vacuum electromagnetic field. In section III, we calculate the evolution of the \textit{$l_1$} norm and relative entropy of coherence for a single-qubit system. In section IV, we study the evolution for a two-qubit system. We summarize and discuss our
conclusions in the last section.

\section{coherence measures and dynamic evolution of a two-level atom system}

In this section we recall the methods to measure coherence in the reference basis which is due to the off-diagonal elements of a density matrix $\rho$ , for instance, the intuitive \textit{$l_1$} norm and the relative entropy of coherence~\cite{Baumgratz}, which can be given by
\begin{equation}\label{norm0}
C_{l_1}(\rho)=\sum_{\substack{i,j\\ i\ne j}}|\rho_{i,j}|\;,
\end{equation}
and
\begin{equation}\label{entropy0}
C_{\rm RE}(\rho)=S(\rho_{\rm diag})-S(\rho)\;,
\end{equation}
respectively. In Eqs. (\ref{norm0}) and (\ref{entropy0}), $\rho$ indicates a state arbitrarily, $S(\rho)=-\textbf{Tr}(\rho\log\rho) $ is the von Neumann entropy, and $\rho_{\rm diag}$ donates a density matrix that delete all off-diagonal elements of $\rho$.

We will discuss the evolution of an open quantum system. The total Hamiltonian of the system with a two-level atom and the fluctuating vacuum electromagnetic field is
$H=H_{s}+H_{f}+H_{I}$, where $H_{s}$ and $H_{f}$ denote the Hamiltonian of the atom and the free electromagnetic field, respectively, and $H_{I}$ represents their interaction. Here we take a two-level atom with Hamiltonian $H_{s}=\frac{1}{2}\hbar\omega_{0}\sigma_{3}$, where $\omega_{0}$ is the energy level spacing, and
 $\sigma_{3}$ is the Pauli matrix. The Hamiltonian representing the interaction between atom and electromagnetic field is
\begin{equation}
H_{I}(\tau)=-e\textbf{r}\cdot\textbf{E}(x(\tau))\;,
\end{equation}
where $e$ is the electron electric charge, $e\textbf{r}$ is the atomic electric dipole moment, $\textbf{E}(x)$ denotes the electric field strength. The initial state of the whole system is given by the total density matrix $\rho_{tot}=\rho(0)\otimes|0\rangle\langle0|$,
where $\rho(0)$ is the reduced density matrix of the two-level atom, and $|0\rangle$ is the vacuum state of the field. For the total system, its equation of motion in Schrodinger picture is
\begin{equation}
\frac{\partial\rho_{tot}(\tau)}{\partial\tau}=-\frac{i}{\hbar}[H,\rho_{tot}(\tau)]\;,
\end{equation}
where $\tau$ is the proper time. In the limit of weak coupling, the evolution of the reduced density matrix $\rho(\tau)$  can be written in the Kossakowski-Lindblad form~\cite{Gorini,Benatti}
\begin{equation}\label{master}
\frac{\partial\rho(\tau)}{\partial\tau}=-\frac{i}{\hbar}[H_{eff},\rho(\tau)]+\mathcal{L}[\rho(\tau)]\;,
\end{equation}
where $H_{eff}$ is the effective Hamiltonian
\begin{eqnarray}\label{master13}
H_{eff}=\frac{1}{2}\hbar\Omega\sigma_{3}=\frac{\hbar}{2}\{\omega_{0}+\frac{i}{2}[\mathcal{K}(-\omega_{0})-\mathcal{K}(\omega_{0})]\}\sigma_{3}\;,\nonumber\\
\end{eqnarray}
and
\begin{eqnarray}
\mathcal{L}[\rho]=\frac{1}{2}\sum_{i,j=1}^{3}a_{ij}[2\sigma_{j}
\rho\sigma_{i}-\sigma_{i}\sigma_{j}\rho-\rho\sigma_{i}\sigma_{j}]\;.\label{master133}
\end{eqnarray}
In Eqs.~(\ref{master13}) and (\ref{master133}), $\Omega$ is the effective energy level-spacing of the atom and $a_{ij}$ is
the coefficients of the Kossakowski matrix which can be expressed as
\begin{equation}
a_{ij}=A\delta_{ij}-iB\epsilon_{ijk}\delta_{k3}-A\delta_{i3}\delta_{j3}\;,
\end{equation}
with
\begin{equation}
A=\frac{1}{4}[\mathcal{G}(\omega_{0})+\mathcal{G}(-\omega_{0})]\;,
\;\;B=\frac{1}{4}[\mathcal{G}(\omega_{0})-\mathcal{G}(-\omega_{0})]\;,
\end{equation}
where $\mathcal{G}(\lambda)$ and $\mathcal{K}(\lambda)$ represent  Fourier and Hilbert transforms respectively, are defined as follows:
\begin{eqnarray}
&&\mathcal{G}(\lambda)=\int_{-\infty}^{\infty}d\triangle\tau e^{i\lambda\triangle\tau}G^+({\triangle\tau})\nonumber \\
&&\mathcal{K}(\lambda)=\frac{P}{\pi i}\int_{-\infty}^{\infty}d\omega\frac{\mathcal{G}(\omega)}{\omega-\lambda}\;.
\end{eqnarray}
In integral the function $G^+({\triangle\tau})$ is given by $G^{+}(x-x')=\frac{e^{2}}{\hbar^{2}}\sum_{i,j=1}^{3}\langle+|r_{i}|-\rangle\langle-|r_{j}|+\rangle\langle0|\textbf{E}_{i}(x)\textbf{E}_{j}(x')|0\rangle$
is the two-point correlation function for electromagnetic field. Here $|+\rangle$, $|-\rangle$
denote the excited state and ground state of the atom.

According to Eq.~(\ref{master}), and assuming a single two-level atom with initial state $|\psi(0)\rangle=\cos\frac{\theta}{2}|1\rangle+e^{i\phi}\sin\frac{\theta}{2}|0\rangle$, its time-dependent reduced density matrix is found to be \begin{widetext}
\begin{equation}  \label{matrix}
\rho(\tau)=\frac{1}{2}
\left(
\begin{array}{cc}
1+\cos\theta e^{-4A\tau}-\frac{B}{A}(1-e^{-4A\tau})& e^{-2A\tau-i(\Omega\tau+\phi)}\sin\theta\\
e^{-2A\tau+i(\Omega\tau+\phi)}\sin\theta&1-\cos\theta e^{-4A\tau}+\frac{B}{A}(1-e^{-4A\tau})
\end{array}
\right).
\end{equation} \end{widetext}

\section{Influence of vacuum fluctuations on quantum coherence for a single-qubit}
Let us now analyze how the initial state and vacuum fluctuations affect the quantum coherence such that the \textit{$l_1$} norm and relative entropy of coherence can be frozen during the dynamic evolution of the atom. We can do it by using the differential of the measures on the evolved state with the noise parameter $q$ that is equal to zero~\cite{thomas}
\begin{equation}\label{condi1}
\partial_{q}C(\rho(q))=0\;,\\\ \forall q\in[0,1],
\end{equation}
where $q=1-e^{-\gamma t}$,  $t$ represents time and $\gamma$ is the decoherence rate~\cite{Nielsen}. In this work we will consider two situations, where the atom interacts with a electromagnetic field without a boundary and a electromagnetic field  with a boundary, respectively.

\subsection{In the case without a reflecting boundary}
In order to calculate the spontaneous emission rates of a static polarizable two-level atom in the electromagnetic field without a boundary, we consider the trajectory of the atom
\begin{eqnarray}\label{trajectory}
t(\tau)=\tau\;, \ \ \ x(\tau)=x_0\;, \ \ \
y(\tau)=y_0\;,\ \ \ z(\tau)=z_0\;.
\end{eqnarray}
According to the trajectory of the atom and using the two-point function of the electromagnetic field in the unbounded space~\cite{Greiner}, we can easily obtain the field correlation function in the frame of the atom
\begin{equation}
G^{+}(x-x')=\sum_{i=1}^{3}\frac{e^{2}c|\langle-|r_{i}|
+\rangle|^{2}}{\hbar\pi^{2}\varepsilon_{0}(c\Delta\tau-i\varepsilon)^{4}}\;,
\end{equation}
and the spontaneous emission rate is given by~\cite{Yu}
\begin{equation}
\gamma_0=\sum_{i=1}^{3}\frac{e^2|\langle -|r_{i}|+\rangle|^2\,\omega_0^3}{3\pi\varepsilon_0\hbar c^3}\;.
\end{equation}
Then we obtain the coefficients of the Kossakowski matrix $a_{ij}$ and the effective level spacing of the atom respectively
\begin{equation}\label{coefficient0}
A^{(0)}=B^{(0)}=\frac{\gamma_0}{4}\;,
\end{equation}
\begin{equation}\label{coefficient1}
\Omega^{(0)}=\omega_0+\frac{\gamma_0}{2\pi\omega_0^3}P\int_0^\infty
d\omega\,\omega^3\bigg(\frac{1}{\omega+\omega_0}-\frac{1}{\omega-\omega_0}\bigg)\;,
\end{equation}
where the superscript $0$ indicates the vacuum fluctuations in the unbounded space.

Using Eqs.~(\ref{matrix}), (\ref{coefficient0}) and (\ref{coefficient1}), the \textit{$l_1$} norm and relative entropy of coherence for the initial single-qubit are found to be
\begin{equation}\label{norm1}
C_{l_{1}}(\rho)^{(0)}=|\sin\theta e^{-\frac{1}{2}\gamma_{0}\tau}|\;,
\end{equation}
and \begin{widetext}
\begin{eqnarray}\label{entropy1}
C_{\rm RE}(\rho)^{(0)}&=&-\frac{1}{2}[1+\cos\theta (1-q)-q]\log_{2}\frac{1}{2}[1+\cos\theta (1-q)-q]\nonumber\\
&-&\frac{1}{2}[1-\cos\theta (1-q)+q]\log_{2}\frac{1}{2}[1-\cos\theta (1-q)+q]\nonumber\\
&+&\frac{1}{2}(1+\sqrt{L})\log_{2}\frac{1}{2}(1+\sqrt{L})+\frac{1}{2}
(1-\sqrt{L})\log_{2}\frac{1}{2}(1-\sqrt{L})\;,
\end{eqnarray}
where $q=1-e^{-\gamma_{0}\tau}$, and $L=(1-\cos^{2}\theta) (1-q)+[\cos\theta (1-q)-q]^{2}$.
\end{widetext}

Now let us analyze under which dynamical conditions the quantum coherence of the single-qubit system is totally unaffected by noise. By applying Eq.~(\ref{norm1}) to Eq.~(\ref{condi1}), the $q$ derivative of the \textit{$l_1$} norm can be calculated as follow
\begin{equation} \label{19}
\frac{|\sin\theta|}{2\sqrt{1-q}}=0\;.
\end{equation}
As the \textit{$l_1$} norm of coherence is dependent of off-diagonal elements of a density matrix, we come to a conclusion that the necessary
and sufficient freezing condition for \textit{$l_1$} norm under the dynamic evolution of the atom is \emph{$\sin\theta=0$} in the initial state.

Similarly, substituting Eq.~(\ref{entropy1}) into Eq.~(\ref{condi1}) we have
\begin{equation}
\log_{2}\bigg[\frac{1-\cos\theta(1-q)+q}{1+\cos\theta(1-q)-q}\bigg]
+\frac{(2q-1)\log_{2}\frac{1-\sqrt{L}}{1+\sqrt{L}}}{2(1+\cos\theta)^{-1}\sqrt{L}}=0
\;.
\end{equation}
From which we find that such a measure is frozen through the dynamic evolution only when $\cos\theta=1$ or $\cos\theta=-1$
(trivial because the initial state is incoherent).

Above discussions tell us that under the dynamic evolution leave coherence invariant \emph{only when} the initial state is incoherent. In other words, the dynamics of quantum
coherence can not be frozen in the studied open quantum system, which is due to the fact that the coherence is affected by the interaction between the two-level atom and the fluctuating vacuum. Unlike the behavior of a single-qubit state~\cite{thomas}, we find that  the \textit{$l_1$} norm of coherence can be frozen for specific initial states, while there is no nontrivial freezing of the relative entropy of coherence.

\subsection{In the case with the presence of a reflecting boundary}
\begin{figure}[htbp]
\centering
\includegraphics[height=1.8in,width=3.4in]{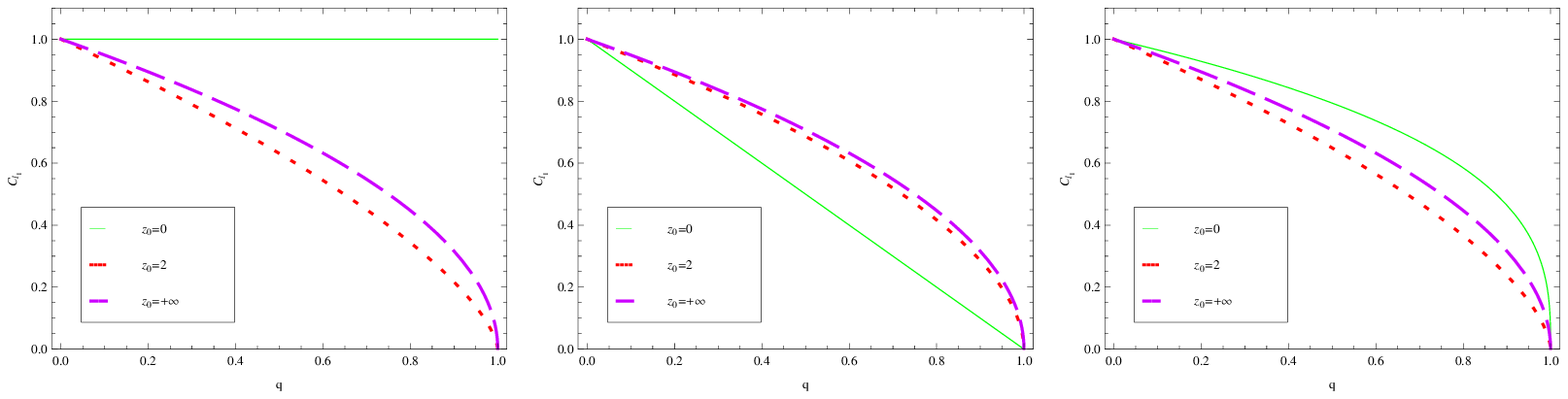}
\caption{ (color online). The \textit{$l_1$} norm of coherence as the function of $q$ and $z_0$ for $(\alpha_x, \alpha_y, \alpha_z)=(1,0,0),(0,0,1),(1/3,1/3,1/3)$ respectively. We set  $\sin\theta=1$ and the values of all the curves are in the unit of $c/\omega_0$.
}\label{f1}
\end{figure}

Now, we calculate the quantum coherence of the two-level atom in the vacuum fluctuations which existing a reflecting boundary. To do so, we consider the case that the atom is placed  at a distance $z_{0}$ from the boundary~\cite{Yu}.
Then the coefficients of the Kossakowski matrix $a_{ij}$ and the effective level spacing of the atom can be calculated as
\begin{equation}
A=B=\frac{\gamma_0}{4}[1-\sum_i\alpha_if_i(\omega_0,z_0)]\;,
\end{equation}
\begin{widetext}
\begin{eqnarray}\label{coefficient2}
\Omega=\omega_0+\frac{\gamma_0}{2\pi\omega_0^3}\,P\int_0^\infty d\omega\,\omega^3\big[1-\sum_i\;\alpha_if_i(\omega_0,z_0)\big]\bigg(\frac{1}{\omega+\omega_0}-\frac{1}{\omega-\omega_0}\bigg)\,\;,
\end{eqnarray}
with
\begin{eqnarray}
&&f_x(\lambda,z_0)=f_y(\lambda,z_0)
=\frac{3c^3}{16\lambda^3z_0^3}\bigg[\frac{2\lambda z_0}{c}\cos\frac{2\lambda z_0}{c}+\bigg(\frac{4\lambda^2z_0^2}{c^2}-1\bigg)\sin\frac{2\lambda z_0}{c}\bigg]\;,\nonumber\\
&&f_z(\lambda,z_0)=\frac{3c^3}{8\lambda^3z_0^3}\bigg[\frac{2\lambda z_0}{c}\cos\frac{2\lambda z_0}{c}-\sin\frac{2\lambda z_0}{c}\bigg]\;,
\end{eqnarray}\end{widetext}
where, $\alpha_i=|\langle-|r_i|+\rangle|^2/|\langle -|{\bf r}|+\rangle|^2$ represents the relative polarizability, which satisfies $\sum_i\alpha_i=1$, and
$f_i(\omega_0,z_0)$ are oscillating functions of distance $z_{0}$ with a position-dependent amplitude. Hereafter, for simplify we  abbreviate $\sum_i\alpha_if_i(\omega_0,z_0)$ as $f(\omega_0,z_0)$. Then the \textit{$l_1$} norm with the presence of a reflecting boundary can be calculated using Eq.~(\ref{norm0})
\begin{equation}\label{norm2}
C_{l_{1}}(\rho)=|\sin\theta e^{-\frac{1}{2}\gamma_{0}[1-f(\omega_{0},z_{0})]\tau}|\;.
\end{equation}
By applying Eq.~(\ref{norm2}) to Eq.~(\ref{condi1}), we obtain the $q$ derivative of the \textit{$l_1$} norm of coherence
\begin{eqnarray}\label{26}
|\sin\theta|[1-f(\omega_{0},z_{0})](1-q)^{-\frac{1}{2}[1+f(\omega_{0},z_{0})]}=0\;.
\end{eqnarray}
Comparing Eq.~(\ref{26}) with (\ref{19}), we can see that  the term $1-f(\omega_{0},z_{0})$ in Eq.~(\ref{26}) in fact decides the dynamical conditions so that the quantum coherence of the single-qubit system is totally unaffected by noise with the presence of a reflecting boundary. Let us now first analyze the limit situation when the atom is placed very far from the boundary, which leads to  $f_i(\omega_0,z_0)\rightarrow0$, and makes this case reduces to that of the unbounded Minkowski vacuum. As the polarizations of the atom are in different directions and behave differently.
To show the properties of the \textit{$l_1$} norm, we plot it as the function of $q$ and $z_0$ for $(\alpha_x, \alpha_y, \alpha_z)=(1,0,0),(0,0,1),(1/3,1/3,1/3)$ in Fig.~(\ref{f1}), which corresponding to the parallel, vertical, isotropic polarization cases respectively.
From Fig. (1) we can see that  due to the presence of the boundary, the vacuum fluctuations is subjected to influenced, and the \textit{$l_1$} norm of  quantum coherence can be frozen under suitable condition.

\emph{Case} \emph{(i) }It is worthy to notice that no matter what direction is the atom polarized in the $xy$ plane, the relations  $\alpha_{z}=0$ and $z_0\rightarrow0$ can be satisfied, which lead to  $f_x(\omega_0,z_0)=f_y(\omega_0,z_0)=1$ in Eq.~(\ref{norm2}). Then $C_{l_{1}}(\rho)$ is independent of the environment, which signifies that the \textit{$l_1$} norm of coherence for the single-qubit system will be frozen when the parallel polarized atom close to the boundary, protecting the important quantum resources of coherence effectively.

\emph{Case} \emph{(ii)} If the atom is polarized in $z$ axis direction, we cam get $\alpha_{x}=\alpha_{y}=0$, which means such a measure is frozen only when trivially the initial state is incoherent, while the decoherence rate decreases slower and the quantum coherence get protected by the presence of the boundary as compared to the unbounded case.

\emph{Case} \emph{(ii)} If the polarization is in an isotropic polarization direction $(\alpha_x=\alpha_y=\alpha_z=\frac{1}{3})$, only trivially when the initial state is already incoherent leads to the coherence be frozen, otherwise the quantum coherence become protected in some degree due to the presence of the boundary.

In the same way, the relative entropy of coherence with the presence of a reflecting boundary will be found using Eq.~(\ref{entropy0}) \begin{widetext}
\begin{eqnarray}\label{entropy2}
C_{\rm RE}(\rho)&=&-\frac{1}{2}\bigg[1+\cos\theta\,(1-q')-q'\bigg]\log_{2}\frac{1}{2}\bigg[1+\cos\theta\,(1-q')-q'\bigg]\nonumber\\
&-&\frac{1}{2}\bigg[1-\cos\theta\,(1-q')+q'\bigg]\log_{2}\frac{1}{2}\bigg[1-\cos\theta\,(1-q')+q'\bigg]\nonumber\\
&+&\frac{1}{2}(1+\sqrt{M})\log_{2}\frac{1}{2}(1+\sqrt{M})+\frac{1}{2}
(1-\sqrt{M})\log_{2}\frac{1}{2}(1-\sqrt{M})\;,
\end{eqnarray} \end{widetext}
where $q'(\tau)=1-e^{-\gamma\tau}$, $\gamma=\gamma_{0}[1-f(\omega_{0},z_{0})]$, and $M=(1-\cos^{2}\theta) (1-q')+[\cos\theta (1-q')-q']^{2}$ respectively.

Substituting Eq.~(\ref{entropy2}) into Eq.~(\ref{condi1}), we obtain the $q$ derivative of the relative entropy of coherence\begin{widetext}
\begin{eqnarray}
\big[1-f(\omega_{0},z_{0})\big](1-q)^{-f(\omega_{0},z_{0})}
\bigg\{\log_{2}\bigg[\frac{1-\cos\theta(1-q')+q'}{1+\cos\theta(1-q')-q'}\bigg]
+\frac{(2q'-1)\log_{2}\frac{1-\sqrt{M}}{1+\sqrt{M}}}{2(1+\cos\theta)^{-1}\sqrt{M}}\bigg\}=0\;.
\end{eqnarray}\end{widetext}

\begin{figure}[htbp]
\centering
\includegraphics[height=1.8in,width=3.4in]{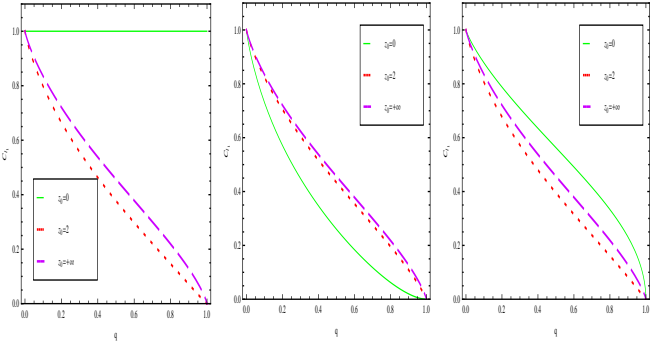}
\caption{ (color online). The relative entropy of coherence is the function of $q$ and $z_0$ for $(\alpha_x, \alpha_y, \alpha_z)=(1,0,0),(0,0,1),(1/3,1/3,1/3)$ respectively. We set  $\sin\theta=1$ and the values of all the curves are in the unit of $c/\omega_0$.
}\label{f2}
\end{figure}

Let us consider the properties of the relative entropy of coherence with the presence of a reflecting boundary. We plot it in Fig.~(\ref{f2}), which is the function of $q$ and $z_0$ for $(\alpha_x, \alpha_y, \alpha_z)=(1,0,0),(0,0,1),(1/3,1/3,1/3)$, corresponding to the parallel, vertical, isotropic polarization cases, respectively. Similar to the \textit{$l_1$} norm, no matter what direction is the atom polarized in the $xy$ plane,  the relative entropy of coherence for the single-qubit system will be frozen when the parallel polarized atom extremely close to the boundary. However, the relative entropy of coherence will be protected in other situation. In a word, due to the modification of the vacuum fluctuations caused by the reflecting boundary, the vacuum fluctuations induced the quantum coherence becomes position dependent.

\section{Influence of vacuum fluctuations on quantum coherence for the two-qubit system}

Now we consider a two-qubit initial state with maximally mixed marginals states, also known as Bell-diagonal ($BD$) states~\cite{Horodecki1,Horodecki2}. The density matrix can be represented as $\rho_{0}=\frac{1}{4}(1_{AB}+\sum_{i=1}^{3}c_{i}\sigma_{i}^{A}\sigma_{i}^{B})$, where $\{\sigma_{i}^{A}\}$ and $\{\sigma_{i}^{B}\}$ denote the Pauli matrices. The initial state can also be expressed by the vector $\vec{c} =(c_{1},c_{2},c_{3})$, which is
\begin{equation}\label{matrix1}
\rho_{0}=\frac{1}{4}
\left(
\begin{array}{cccc}
1+c_{3}& 0&0&c_{1}-c_{2}\\
0& 1-c_{3}&c_{1}+c_{2}&0\\
0&c_{1}+c_{2}&1-c_{3}&0\\
c_{1}-c_{2}&0&0&1+c_{3}
\end{array}
\right).
\end{equation}
According to Eq.~(\ref{master}),  the corresponding time-dependent density matrix is\begin{widetext}
\begin{eqnarray}
\rho_{AB}(\tau)=\frac{1}{4}\{\sigma_{0}^{A}\otimes\sigma_{0}^{B}+c_{1}e^{-2A\tau}\cos(\Omega\tau)\sigma_{1}^{A}\otimes\sigma_{1}^{B}
+c_{2}e^{-2A\tau}\cos(\Omega\tau)\sigma_{2}^{A}\otimes\sigma_{2}^{B}+c_{3}e^{-2A\tau}\sigma_{3}^{A}\otimes\sigma_{3}^{B}\nonumber\\
-c_{2}e^{-2A\tau}\sin(\Omega\tau)\sigma_{1}^{A}\otimes\sigma_{2}^{B}+c_{1}e^{-2A\tau}\sin(\Omega\tau)\sigma_{2}^{A}\otimes\sigma_{1}^{B}
-\frac{B}{A}(1-e^{-4A\tau})\sigma_{3}^{A}\otimes\sigma_{0}^{B}\}\;.
\end{eqnarray}\end{widetext}

Comparing to the analysis of the single-qubit system above, we have to consider the influence of the initial state and vacuum fluctuations on quantum coherence of the  two-qubit system. We also interested in under which dynamical conditions the coherence is totally unaffected by noise for the two-qubit system. In the following  we adopt a bona fide distance-based measure of quantum coherence~\cite{thomas}
\begin{equation}\label{condi2}
C_{D}(\rho(q))=C_{D}(\rho_{0})\;,
\end{equation}
which is frozen forever for any $q\in[0,1]$, or equivalently frozen for any $\tau$~\cite{Nielsen}, meaning that the quantum coherence of the system is independent of evolution.

Substituting Eq.~(\ref{matrix1}) into Eq.~(\ref{norm0}) and Eq.~(\ref{entropy0}), the \textit{$l_1$} norm and relative entropy of the initial two-qubit state can be easily obtained as
\begin{equation}\label{norm3}
C_{l_{1}}(\rho_{0})=\frac{1}{2}\big(|c_{1}+c_{2}|+|c_{1}-c_{2}|\big)\;,
\end{equation}
and\begin{widetext}
\begin{eqnarray}\label{entropy3}
C_{\rm RE}(\rho_{0})&=&\frac{1}{4}\bigg\{-\big[2(1+c_{3})\log_{2}\frac{1}{4}(1+c_{3})+2(1-c_{3})\log_{2}\frac{1}{4}(1-c_{3})\big]\nonumber\\
&+&[1+c_{3}+(c_{1}-c_{2})]\log_{2}\frac{1}{4}[1+c_{3}+(c_{1}-c_{2})]+[1+c_{3}-(c_{1}-c_{2})]\log_{2}\frac{1}{4}[1+c_{3}-(c_{1}-c_{2})]\nonumber\\
&+&[1-c_{3}+(c_{1}+c_{2})]\log_{2}\frac{1}{4}[1-c_{3}+(c_{1}+c_{2})]+[1-c_{3}-(c_{1}+c_{2})]\log_{2}\frac{1}{4}[1-c_{3}-(c_{1}+c_{2})]\bigg\}\;.\nonumber\\
\end{eqnarray}\end{widetext}

\subsection{In the case without a reflecting boundary}
Let us  consider the case of the vacuum fluctuations without a boundary. Noting that the electric-field two-point functions which without a boundary~\cite{Greiner} and the trajectory Eq.~(\ref{trajectory}) are considered above. As a result, the \textit{$l_1$} norm and relative entropy under the evolution are found to be
\begin{equation}\label{norm4}
C_{l_{1}}(\rho_{AB})^{(0)}=\frac{1}{2}e^{-\frac{\gamma_{0}}{2}\tau}\big(|c_{1}+c_{2}|+|c_{1}-c_{2}|\big)\;.
\end{equation}
and\begin{widetext}
\begin{eqnarray}\label{entropy4}
C_{\rm RE}(\rho_{AB})^{(0)}&=&
\frac{1}{4}\bigg\{-\big[1+c_{3}(1-q)-q\big]\log_{2}\frac{1}{4}[1+c_{3}(1-q)-q\big]-\big[1+c_{3}(1-q)+q\big]\log_{2}\frac{1}{4}[1+c_{3}(1-q)+q\big]\nonumber\\
&-&\big[1-c_{3}(1-q)-q\big]\log_{2}\frac{1}{4}[1-c_{3}(1-q)-q\big]-\big[1-c_{3}(1-q)+q\big]\log_{2}\frac{1}{4}[1-c_{3}(1-q)+q\big]\nonumber\\
&+&(1+c_{3}(1-q)-N)\log_{2}\frac{1}{4}\big[1+c_{3}(1-q)-N\big]+(1+c_{3}(1-q)+N)\log_{2}\frac{1}{4}\big[1+c_{3}(1-q)+N\big]\nonumber\\
&+&(1-c_{3}(1-q)-N)\log_{2}\frac{1}{4}\big
[1-c_{3}(1-q)-N\big]+(1-c_{3}(1-q)+N)\log_{2}\frac{1}{4}\big
[1-c_{3}(1-q)+N\big]\bigg\}\;,\nonumber\\
\end{eqnarray}\end{widetext}
where $N=\sqrt{(1-q)(c_{1}+c_{2})^{2}+q^{2}}$, and the superscript $0$ indicates the vacuum fluctuations in the unbounded space. Now let us analyze under which dynamical conditions the quantum coherence of the two-qubit system is totally unaffected by noise. By applying Eqs.~(\ref{norm3})$-$(\ref{entropy4}) to Eq.~(\ref{condi2}),
we  find that the coherence of $BD$ states will be frozen for any bona fide distance-based measure only when
\begin{equation}
c_{1}=c_{2}=0\;.
\end{equation}
Therefore, there is no exist trivial freezing of coherence for the dynamics of the two-qubit system, which means that the quantum coherence of $BD$ states must be affected by noise with time. 

\subsection{In the case with the presence of a reflecting boundary}
However, since vacuum fluctuations will be modified for the present of boundary in the vacuum, it is necessary to examine how the vacuum fluctuations affect the quantum coherence. According to the electric-field two-point functions  with the presence of a boundary~\cite{Yu} and the trajectory Eq.~(\ref{trajectory}),  the \textit{$l_1$} norm and relative entropy of the coherence under the evolution can be easily obtained
\begin{equation}\label{norm5}
C_{l_{1}}(\rho_{AB})=\frac{1}{2}e^{-\frac{\gamma_{0}}{2}[1-f(\omega_{0},z_{0})]\tau}\big(|c_{1}+c_{2}|+|c_{1}-c_{2}|\big)\;,
\end{equation}
and\begin{widetext}
\begin{eqnarray}\label{entropy5}
C_{\rm RE}(\rho_{AB})&=&
\frac{1}{4}\bigg\{-\big[1+c_{3}(1-q')-q'\big]\log_{2}\frac{1}{4}[1+c_{3}(1-q')-q'\big]-\big[1+c_{3}(1-q')+q'\big]\log_{2}\frac{1}{4}[1+c_{3}(1-q')+q'\big]\nonumber\\
&-&\big[1-c_{3}(1-q')-q'\big]\log_{2}\frac{1}{4}[1-c_{3}(1-q')-q'\big]-\big[1-c_{3}(1-q')+q'\big]\log_{2}\frac{1}{4}[1-c_{3}(1-q')+q'\big]\nonumber\\
&+&(1+c_{3}(1-q')-S)\log_{2}\frac{1}{4}\big[1+c_{3}(1-q')-S\big]+(1+c_{3}(1-q')+S)\log_{2}\frac{1}{4}\big[1+c_{3}(1-q')+S\big]\nonumber\\
&+&(1-c_{3}(1-q')-S)\log_{2}\frac{1}{4}\big[1-c_{3}(1-q')-S\big]+(1-c_{3}(1-q')+S)\log_{2}\frac{1}{4}\big[1-c_{3}(1-q')+S\big]\bigg\}\;,\nonumber\\
\end{eqnarray}\end{widetext}
where $S=\sqrt{(1-q')(c_{1}+c_{2})^{2}+q'^{2}}$. According to Eq.~(\ref{condi2}), the dynamical condition of the quantum coherence of the two-qubit system is determined by the term $1-f(\omega_{0},z_{0})$ when a boundary is present. Similar to the analysis of the case of the single-qubit system, we can see that whatever direction the polarization is in the $xy$ plane one has $\alpha_{z}=0$, resulting in the factor $1-f(\omega_{0},z_{0})=0$. This dynamical conditions for freezing the quantum coherence can be fulfilled if the atom is transversely polarized and very close to the boundary, which means that the coherence of a two-qubit system can be frozen. Otherwise, the coherence must be affected by noise under the dynamic evolution in open quantum systems and can only be protected in some degree.

\section{conclusion}
In the framework of open quantum systems, we have investigated the dynamic evolution of a single-qubit system and a two-qubit system when the atom is coupling with the bath of fluctuating vacuum electromagnetic field.
Two different situations, the atom
interacts with an electromagnetic field
without or with the presence of a reflecting boundary, have been studied.
For a single-qubit system, we find that for the case of without a boundary, the quantum coherence will be destroyed  due to the interaction between the atom and the electromagnetic field. However, with the presence of a boundary, the quantum coherence becomes dependent on position and atomic polarization. The dynamical conditions for the insusceptible of coherence are fulfilled \emph{only when} the atom is close to the boundary and is transversely polarizable. Otherwise, the quantum coherence can only be protected in some degree in other polarizable direction. Similarly, for a two-qubit system, the vacuum fluctuations always affect the quantum coherence without a boundary. With the presence of a boundary, the quantum coherence of two-qubit states with maximally mixed marginals is shielded from the influence of the vacuum fluctuations of the electromagnetic field when the atom is close to the boundary and is transversely polarizable.

\begin{acknowledgments}
This work is supported by the  National Natural Science Foundation
of China under Grant Nos. 11475061 and 11305058;  the SRFDP under Grant No.
20114306110003; the Open Project Program of State Key Laboratory of
Theoretical Physics, Institute of Theoretical Physics, Chinese
Academy of Sciences, China (No.Y5KF161CJ1); and
the  Postdoctoral Science Foundation of China under Grant No. 2015T80146.

\end{acknowledgments}

\end{document}